\documentclass[alpha-refs]{wiley-article}

\usepackage{siunitx}
\usepackage{booktabs}
\usepackage{threeparttable}
\usepackage{tabularx}
\usepackage{array}
\usepackage{multirow}
\usepackage{rotating}
\usepackage{amsmath}
\usepackage{microtype}
\usepackage{xcolor}
\usepackage{enumitem}
\makeatletter
\renewcommand{\marginpar}[2][]{ }
\newcommand{\fullwidthauthormetadata}{%
  \par\noindent
  \begin{minipage}{\textwidth}
    \AB@affillist\par
    \vspace{6pt}
    {\footnotesize\textbf{Correspondence:} \@corraddress\par
    \textbf{Email:} \@corremail\par}
  \end{minipage}
  \par\vspace{12pt}
}

\renewenvironment{abstract}{%
  \par\noindent
  \begin{mdframed}[
    font=\fontsize{9pt}{15pt}\selectfont,
    topline=false,bottomline=false,
    leftline=true,rightline=false,
    backgroundcolor=black!10,
    middlelinewidth=6pt,middlelinecolor=white,
    outerlinewidth=0.5pt]
}{%
  \end{mdframed}
  \par
}

\providecommand{\keywords}[1]{%
  \vspace{8pt}%
  \noindent{\bfseries\fontsize{7pt}{13pt}\selectfont KEYWORDS\par}%
  \noindent{\raggedright\fontsize{8pt}{13pt}\selectfont #1\par}%
}
\makeatother

\papertype{Original Article}
\paperfield{Sociobiology and Behaviour}

\title{How long versus when: worker mortality synchrony is widespread but
  decoupled from body size in Australian ants}

\abbrevs{%
  ICC, intra-class correlation coefficient;
  CrI, credible interval;
  WLS, weighted least squares;
  SE, standard error;
  SD, standard deviation;
  CI, confidence interval%
}

\author[1]{Alana Moscardi}
\author[2]{Rafael da Silva}
\author[3]{Gleycon Silva}

\affil[1]{Graduate Programme in Ecology, Instituto Nacional de Pesquisas
  da Amaz\^{o}nia (INPA), Manaus, Brazil}
\affil[2]{Graduate Programme in Data Science, Eastern University,
  St.\ Davids, PA, USA}
\affil[3]{Graduate Programme in Ecology, Instituto Nacional de Pesquisas
  da Amaz\^{o}nia (INPA), Manaus, Brazil}

\corraddress{Alana Moscardi, Graduate Programme in Ecology, Instituto
  Nacional de Pesquisas da Amaz\^{o}nia (INPA), Manaus, AM, Brazil}
\corremail{alanamoscardi1@gmail.com}

\runningauthor{Moscardi et al.}
\fundinginfo{[To be completed before submission]}

\begin{document}
\begin{frontmatter}
\maketitle
\fullwidthauthormetadata

\begin{abstract}
{\small
\begin{enumerate}[leftmargin=1.4em, itemsep=2pt, topsep=1pt, parsep=0pt]
  \item Colony demography models treat worker deaths as independent events, yet
    shared environmental exposures can synchronise mortality within species;
    whether such synchrony exists and what drives it remain open questions.

  \item Using a 15-month field dataset of 18 Australian ant species and 106
    cohorts \citep{Riskas2026}, we quantified intra-specific mortality synchrony
    with the intra-class correlation coefficient (ICC) and tested body size and
    temperature as potential correlates of synchrony.

  \item \textbf{Synchrony is widespread.} Credible intervals excluded zero in 9
    of 18 species (assemblage mean ICC\,=\,$0.21 \pm 0.09$); on average,
    ${\sim}21\,\%$ of daily mortality variance reflects shared within-species
    exposure rather than individual stochasticity.

  \item \textbf{Body size does not buffer synchrony.} Contrary to a buffering
    prediction, body mass was uncorrelated with ICC (Spearman $\rho = +0.10$,
    $p = 0.69$; WLS $\beta = +0.05$, $p = 0.35$), decoupling \emph{how long}
    workers live from \emph{when} they die together.

  \item \textbf{Temperature is a suggestive, lagged correlate.} Maximum temperature
    preceded mortality pulses by 1--7 days in 13 of 15 species (mean peak
    $r = 0.16$), suggesting a consistent but modest thermal signal detectable
    only at assemblage level (no species survived Bonferroni correction,
    $\alpha_{\rm adj} = 0.0033$); a sensitivity analysis using mean $r$ across
    lags 1--7 days rather than the peak yielded $\bar{r} = -0.03 \pm 0.08$ (SD),
    $p = 0.93$, indicating that the assemblage-level signal is sensitive to
    peak-lag selection; yet ICC was unrelated to
    temperature-coupling strength ($\rho = -0.004$, $p = 0.99$).

  \item Mortality synchrony is thus widespread, size-independent, and only
    partly thermal; species-specific thermal tolerance, not body mass, is the
    more promising candidate predictor.
\end{enumerate}
}


\end{abstract}
\end{frontmatter}

\section{Introduction}
\label{sec:intro}

\subsection{An overlooked assumption: independent worker mortality}

During a heat event, ant workers across foraging parties and brood-tending
teams can die in concentrated pulses --- whole cohorts lost within a few days
rather than the steady attrition that independent survival would predict.
Yet colony demography models rarely account for this; many treat
worker deaths as statistically independent events
\citep{Chouvenc2022, Bourke1999}.  Workers of the same species share foraging
trails, nest microhabitats, and collective thermoregulation, and face the same
weather, predators, and pathogens.  Any of these shared exposures can
synchronise cohort mortality --- temporal clustering we term
\emph{intra-specific mortality synchrony}.  Synchrony matters because it
inflates the \emph{variance}, not just the mean, of worker availability:
deaths that arrive in pulses can sharply reduce the effective worker force at
critical moments (brood care, colony defence, reproductive flights) in a way
that gradual, independent attrition would not.

We organise the study around three questions, each resolved by a falsifiable
claim (Table~\ref{tab:roadmap}).

\begin{table}[!htb]
\caption{Analytical roadmap: questions, claims, internal hypotheses, and
  outcomes.  Internal labels (H1a--H1d) cross-reference the Methods, code,
  and supplementary material.
  \label{tab:roadmap}}
\begin{threeparttable}
\small
\begin{tabularx}{\textwidth}{p{3.0cm} p{3.2cm} p{4.0cm} X}
\toprule
\textbf{Question} & \textbf{Claim} &
\textbf{Analysis (internal label)} & \textbf{Outcome}\\
\midrule
Q1.\ Is mortality synchronised?
  & Claim 1.\ Synchrony is widespread.
  & ICC $> 0$ per species (H1a) & $\checkmark$ 9/18 CrI$_{\rm lo} > 0.01$; mean ICC\,=\,0.21\\
  & & Technical check: Wilcoxon & {\footnotesize $W = 0$, $r = 1.00$ (see note\,a)}\\
\midrule
Q2.\ Does body size predict synchrony?
  & Claim 2.\ Size does not predict synchrony.
  & Spearman $\rho$(mass, ICC) (H1b) & $\circ$ $\rho = +0.10$, $p = 0.69$\\
  & & WLS ICC $\sim$ mass + subfamily (H1b) & $\circ$ $\beta = +0.05$, $p = 0.35$\\
  & & Subfamily rank ICC (H1d) & $\triangleright$ $\rho = -0.30$, $p = 0.62$\\
\midrule
Q3.\ Does temperature correlate with synchrony?
  & Claim 3.\ Temperature is a suggestive, lagged correlate (sensitive to peak-selection; see Table~S2).
  & Mortality lags temp 1--7\,d (H1c) & $\triangleright$ $W = 117$, $p = 0.0002$; no species survives Bonferroni\\
  & & Sensitivity: mean $r$ lags 1--7\,d & $\circ$ $\bar{r} = -0.03$, $p = 0.93$\\
  & & ICC $\not\leftrightarrow$ temp coupling & $\circ$ $\rho = -0.004$, $p = 0.99$\\
\bottomrule
\end{tabularx}
\begin{tablenotes}
\footnotesize
\item[] $\checkmark$ Supported. \quad $\circ$ Null result. \quad
  $\triangleright$ Directional (consistent, not conclusive).
\item[] Internal labels H1a--H1d cross-reference the Methods and
  supplementary material: H1a\,=\,ICC positivity; H1b\,=\,body-size
  regression; H1c\,=\,temperature cross-correlation; H1d\,=\,subfamily
  rank test.
\item[] For a non-negative quantity estimated under Bayesian shrinkage,
  the Wilcoxon test is expected to be significant irrespective of true ICC
  magnitude; it confirms that all posterior means are positive but does not
  constitute evidence of synchrony strength.
\end{tablenotes}
\end{threeparttable}
\end{table}

\subsection{Study system}

The dataset derives from the La~Trobe Wildlife Sanctuary, a 28-ha reserve on
La~Trobe University's Melbourne campus, Victoria, in temperate south-eastern
Australia (${\sim}37.7^\circ$\,S, $145.1^\circ$\,E).  The site has a mesic,
temperate climate (mean annual minimum and maximum air temperatures of
9.7\,\textdegree C and 20.2\,\textdegree C; ${\sim}656$\,mm annual rainfall)
and supports a diverse ground-foraging ant assemblage.  Across the study,
species-level mean daily maximum temperature ranged from 21.0 to
24.4\,\textdegree C, with individual daily maxima spanning
13.1--40.5\,\textdegree C --- occasional heat events well above the
seasonal norm and the primary synchronising force tested here.

\textbf{Q1 --- Is worker mortality synchronised within species, and how
strong is it?}  Synchrony can be quantified by the intra-class correlation
coefficient (ICC): the proportion of total mortality variance attributable to
shared, within-species (across-cohort) effects rather than to individual
stochasticity.  An ICC indistinguishable from zero would vindicate the
independence assumption; a positive ICC would reveal a structured, shared
component to mortality \citep{Nakagawa2026, McCune2025}.

\textbf{Q2 --- Does body size, the predictor of mean longevity, also predict
synchrony?}  Body mass predicts mean worker longevity across this assemblage
\citep{Riskas2026}: larger workers survive at higher daily rates, plausibly
through lower mass-specific metabolic rates, reduced oxidative damage, and
greater energetic reserves \citep{Ricklefs2002, Heinze2008, Brace2017,
Kramer2013}.  Under a pace-of-life logic, one might predict that
larger-bodied species, being more robust, are also better buffered against
extrinsic synchronising shocks and so show \emph{lower} ICC.  This generates
a clear, directional test: if size buffers synchrony, ICC should decline with
body mass.

\textbf{Q3 --- Does temperature correlate with synchronised mortality, and how
much does it explain?}  Ambient temperature is a ubiquitous extrinsic force,
and because all cohorts of a species share the same local thermal environment,
heat events are a natural candidate driver of synchronised mortality pulses.
The physiological effect of heat is not instantaneous: thermal stress
initiates cumulative damage --- oxidative injury, protein denaturation,
and other heat-induced stressors --- that may manifest as elevated
mortality days after exposure
\citep{Colinet2015, Sales2021}.  We therefore test whether species-level
daily mortality cross-correlates with maximum temperature at a \emph{positive
lag} (temperature leading mortality), and whether the strength of that
coupling explains variation in ICC.

\section{Materials and Methods}
\label{sec:methods}

The three claims map onto a set of variance-partitioning and time-series
analyses; internal labels (H1a--H1d; Table~\ref{tab:roadmap}) are used here
and in the tables for precise cross-reference.

\subsection{Dataset}

We used the field survival dataset of \citet{Riskas2026}, archived at Dryad
(doi: 10.5061/dryad.j0zpc86s4), comprising 2,363 cohort-day observations for
106 cohorts across 39 colonies and 18 species (February 2022 -- May 2023),
with daily worker counts, maximum temperature, and colony-level traits.
Cohorts were included if they had at least 10 workers alive on day 1 and at
least one death event recorded; cohorts with no mortality throughout the
observation window were excluded from the ICC estimation in the source model
\citep{Riskas2026} and are therefore absent from the present dataset.

\textbf{ICC estimates.}  We used the published posterior means and 95\,\%
Bayesian credible intervals (CrI) for species-level ICC from the hierarchical
survival model of \citet{Riskas2026}, available in the archived dataset
(doi: 10.5061/dryad.j0zpc86s4).  These
quantify the proportion of variance in daily survival probability
attributable to cohort-level (shared, within-species) effects, as opposed to
individual stochasticity.

\textbf{Daily mortality rates.}  Species-level daily mortality rate was
computed as the mean, across cohorts active on a given day, of the daily
cohort mortality rate (deaths / workers alive).  Maximum daily temperature
(\texttt{maxt}) was available for every observation day.

\textbf{Worker body mass.}  Log\textsubscript{10}-transformed mean worker
body mass (log\textsubscript{10} mg) was taken from the species-level trait
table archived alongside the survival data in \citet{Riskas2026}
(Dryad doi: 10.5061/dryad.j0zpc86s4), based on field-collected workers
from the same study colonies.

\subsection{Does mortality synchrony exist? (Claim 1)}

We assessed H1a in two complementary ways; the per-species credible-interval
classification is treated as the primary assessment, with the assemblage-level
Wilcoxon test as a technical confirmation:

\begin{enumerate}
  \item \textbf{Per-species credible-interval classification (primary):}
    Strong (CrI$_{\rm lo} > 0.05$), Moderate (CrI$_{\rm lo} > 0.01$),
    Weak (CrI$_{\rm lo} > 0.00$), or None (CrI includes 0).
  \item \textbf{Assemblage-level test:} one-sample Wilcoxon signed-rank test
    on the 18 ICC posterior means against a null of 0 (one-sided, greater).
    Effect size reported as $r = W / (n(n+1)/2)$ \citep{Kerby2014}.
\end{enumerate}

\subsection{Body size as a predictor of synchrony (Claim 2)}

\begin{itemize}
  \item \textbf{H1b} (species level): Spearman rank correlation between
    log\textsubscript{10} worker body mass and ICC posterior mean; and
    weighted least squares (WLS) regression
    \texttt{ICC\,$\sim$\,logw\_c\,+\,C(subfamily)}, weighted by
    $1/\text{CrI\_width}^2$ (downweighting uncertain ICC estimates;
    in practice weights vary little, as CrI widths range from 0.47 to 0.54
    across species), with
    continuous predictors grand-mean centred.  Both the subfamily-controlled
    model and a simple \texttt{ICC\,$\sim$\,logw\_c} model are reported.
  \item \textbf{H1d} (subfamily level, $N = 5$): Spearman correlation
    between subfamily mean body-mass rank and subfamily mean ICC rank.
\end{itemize}

Given wide credible intervals on ICC (mean CrI width\,$\approx$\,0.50),
the regression has limited power to detect moderate associations.  We
interpret negative or null results as evidence against the specific
buffering prediction (ICC declining with mass), not as proof of zero
association.  Including subfamily as a fixed effect (five levels) in the WLS
provides a partial control for phylogenetic non-independence given the
congeneric clustering in the assemblage, though it does not substitute for
a formal phylogenetic analysis.  As a formal robustness check, we fitted
phylogenetic generalised least squares models \citep[PGLS;][]{Felsenstein1985}
for both ICC\,$\sim$\,body size ($N = 18$) and ICC\,$\sim$\,$r_{\rm peak}$
($N = 15$), using a Brownian-motion covariance matrix derived from the
maximum-clade-credibility tree of \citet{Riskas2026}.  Pagel's $\lambda$
was estimated by maximum likelihood.

\subsection{Temperature as a correlate of synchronised mortality (Claim 3)}

For each species with $\geq 20$ observation days ($N = 15$ species), we
computed Pearson cross-correlation between the species-level daily mortality
time series and maximum daily temperature at lags $-3$ to $+14$ days,
recording the lag-0 (contemporaneous) correlation and the peak correlation
within the 1--7 day window; ties in peak $r$ were broken by selecting the
smallest lag.  No detrending or normalisation was applied prior to
cross-correlation; all daily mortality values and temperature values are used
as computed, with days absent from the record omitted pair-wise at each lag.
Three species were excluded for insufficient
coverage: \textit{Camponotus claripesA} ($n = 18$ days),
\textit{Crematogaster laeviceps} ($n = 9$ days), and
\textit{Iridomyrmex notialis} ($n = 15$ days).  We tested H1c at the
assemblage level by a one-sample Wilcoxon signed-rank test on the 15 peak-lag
$r$ values (one-sided, greater), and tested whether ICC predicts
temperature-coupling strength by Spearman correlation between species ICC and
peak $r$.

We note that recording the peak $r$ within a 7-lag window upwardly biases
the expected cross-correlation under the null, as it selects the maximum
of seven correlated test statistics.  The assemblage-level Wilcoxon test
on these peak values is best interpreted as a test of the consistent
\emph{direction} of temperature coupling across species rather than a
precise estimate of coupling magnitude.  As a pre-specified sensitivity
analysis to address this bias, we additionally computed the unweighted mean
Pearson $r$ across all lags within the 1--7 day window for each species and
applied the same one-sided Wilcoxon test to the assemblage distribution of
these mean values; results are reported in Table~S2 (Supplementary Material).

Both the daily mortality rate and the temperature series carry seasonal
trends: temperature peaks in summer and cohort mortality rates vary over the
15-month observation window.  We analysed raw (non-detrended) daily series,
so any shared seasonal trend inflates the contemporaneous (lag 0)
cross-correlation; the 1--7 day lag window targets sub-seasonal, event-scale
coupling and is therefore less susceptible to seasonal confounding than the
lag-0 estimate.  Positive autocorrelation in both series reduces effective
degrees of freedom and renders per-species $p$-values anticonservative;
we corrected for this at the individual-species level by computing an effective
sample size $N_{\rm eff}$ using the Pyper--Peterman formula
\citep{PyperPeterman1998} applied to the paired series at each lag
(see Table~\ref{tab:xcorr}, note\,b; PP-corrected $p$-values are also provided
in the supplementary data files).
The assemblage-level Wilcoxon test on the 15 peak-$r$ values remains valid
as a test of consistent direction across species regardless of autocorrelation
(see Limitations).

All analyses used Python~3.13 (\texttt{scipy}~1.15.3, \texttt{statsmodels}~0.14.4,
\texttt{pandas}~2.2.3, \texttt{lifelines}~0.30.1).  Analysis scripts are available at
\url{https://github.com/rafa-rodriguess/australian_ants_how_long_versus_when_worker_mortalit}.

\section{Results}
\label{sec:results}

\subsection{Claim 1: Mortality synchrony is widespread}

Worker mortality shows a detectable shared component across the assemblage.
Credible intervals excluded zero (CrI$_{\rm lo} > 0.01$) in 9 of 18 species
(Table~\ref{tab:icc_summary}), indicating that roughly half the assemblage shows
mortality synchrony distinguishable from independence.  The assemblage mean ICC
was $0.21 \pm 0.09$ (SD), ranging from 0.103 (\textit{Melophorus turneri}) to
0.400 (\textit{Chelaner cinctum}).  On average, ${\sim}21\,\%$ of daily
mortality variance is attributable to shared, within-species exposure rather
than to individual stochasticity.

As a technical confirmation that all posterior means are positive, a one-sample
Wilcoxon signed-rank test against zero yielded $W = 0$, $p < 0.001$,
$r = 1.00$; however, for a non-negative quantity estimated under Bayesian
shrinkage this outcome is expected regardless of true ICC magnitude,
and the per-species CrI evidence above
provides the primary assessment of synchrony strength.

\begin{figure}[!htb]
\centering
\includegraphics[width=0.65\textwidth]{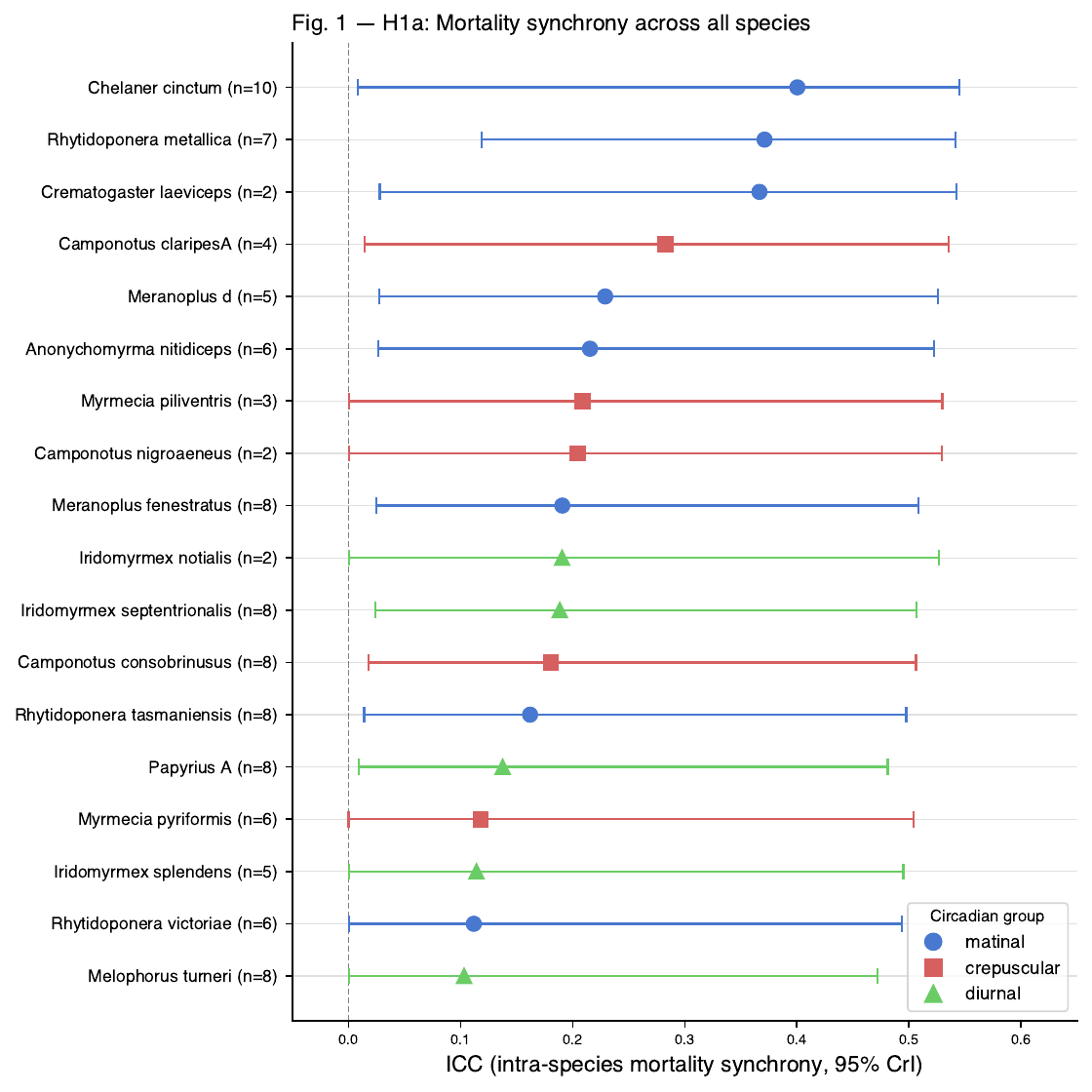}
\caption{\textbf{Per-species ICC with 95\,\% Bayesian credible intervals
  (Claim~1, H1a).}  Species ordered by ICC posterior mean (lowest to
  highest).  Circadian activity group is encoded by both colour and point
  shape (matinal\,=\,blue circle; crepuscular\,=\,red square;
  diurnal\,=\,green triangle) so that groups remain distinguishable in
  greyscale and for colour-vision-deficient readers.  Numbers in parentheses
  on the $y$-axis give the cohort count ($n$) per species.  CrI: 95\,\%
  Bayesian credible interval from the hierarchical survival model of
  \citet{Riskas2026}.  The vertical dashed line marks ICC\,=\,0 (independence
  null).  All 18 species have ICC posterior means above zero ($W = 0$,
  $p < 0.001$, $r = 1.00$); credible intervals exclude zero
  (CrI$_{\rm lo} > 0.01$) in 9 of 18 species.  The two highest-ICC species
  --- \textit{Chelaner cinctum} (small Myrmicinae) and \textit{Rhytidoponera
  metallica} (large Ectatomminae) --- are labelled to illustrate that
  synchrony is not confined to a single morphological guild.
  \label{fig:icc_per_species}}
\end{figure}

Evidence strength varies markedly across species: one met the Strong
criterion (CrI$_{\rm lo} > 0.05$: \textit{Rhytidoponera metallica},
ICC\,=\,0.371), eight Moderate, and nine Weak
(Table~\ref{tab:h1a_test}, Fig.~\ref{fig:icc_per_species}).  The two
highest-ICC species are ecologically disparate ---
\textit{Chelaner cinctum}, a small Myrmicinae (ICC\,=\,0.400), and
\textit{Rhytidoponera metallica}, a large Ectatomminae (ICC\,=\,0.371) ---
indicating that synchrony is not confined to a single morphological or
ecological guild.

\begin{sidewaystable}[p]
\caption{Species-level summary statistics.  Species ordered by ICC posterior
  mean (descending).  CrI: 95\,\% Bayesian credible interval.  Circadian
  group: M\,=\,matinal, D\,=\,diurnal, C\,=\,crepuscular.
  Cross-correlation results ($r_{\rm peak}$ and optimal lag) for the 15
  eligible species are in Table~\ref{tab:xcorr}.
  \label{tab:icc_summary}}
\begin{threeparttable}
\footnotesize
\begin{tabular}{llrrrrrr}
\toprule
\textbf{Species} & \textbf{Subfamily} &
\textbf{Niche} &
\multicolumn{1}{c}{\textbf{log\textsubscript{10}}} &
\multicolumn{1}{c}{\textbf{Cohorts}} &
\multicolumn{1}{c}{\textbf{ICC}} &
\multicolumn{1}{c}{\textbf{CrI}} &
\multicolumn{1}{c}{\textbf{Maxt}}\\
 & & &
\textbf{mass (mg)} & & \textbf{mean} & \textbf{[lo, hi]} &
\textbf{(\textdegree C)}\\
\midrule
\textit{Chelaner cinctum}             & Myrmicinae     & M & $-$0.745 & 10 & 0.400 & [0.008, 0.545] & 22.4\\
\textit{Rhytidoponera metallica}      & Ectatomminae   & M &  $+$0.539 &  7 & 0.371 & [0.119, 0.542] & 21.9\\
\textit{Crematogaster laeviceps}      & Myrmicinae     & M & $-$0.446 &  2 & 0.367 & [0.028, 0.543] & 23.5\tnote{$\dagger$}\\
\textit{Camponotus claripesA}         & Formicinae     & C &  $+$0.041 &  4 & 0.283 & [0.015, 0.536] & 23.8\tnote{$\dagger$}\\
\textit{Meranoplus d}                 & Myrmicinae     & M & $-$0.796 &  5 & 0.229 & [0.027, 0.526] & 22.2\\
\textit{Anonychomyrma nitidiceps}     & Dolichoderinae & M & $-$0.320 &  6 & 0.215 & [0.026, 0.523] & 23.6\\
\textit{Myrmecia piliventris}         & Myrmeciinae    & C &  $+$0.980 &  3 & 0.209 & [0.001, 0.530] & 22.5\\
\textit{Camponotus nigroaeneus}       & Formicinae     & C & $-$0.094 &  2 & 0.205 & [0.000, 0.530] & 21.0\\
\textit{Iridomyrmex notialis}         & Dolichoderinae & D & $-$0.955 &  2 & 0.191 & [0.000, 0.527] & 22.7\tnote{$\dagger$}\\
\textit{Meranoplus fenestratus}       & Myrmicinae     & M &  $+$0.049 &  8 & 0.191 & [0.025, 0.509] & 23.1\\
\textit{Iridomyrmex septentrionalis}  & Dolichoderinae & D & $-$0.399 &  8 & 0.188 & [0.024, 0.507] & 24.4\\
\textit{Camponotus consobrinusus}     & Formicinae     & C &  $+$0.944 &  8 & 0.181 & [0.018, 0.506] & 23.3\\
\textit{Rhytidoponera tasmaniensis}   & Ectatomminae   & M &  $+$0.136 &  8 & 0.162 & [0.014, 0.498] & 21.1\\
\textit{Papyrius} A                   & Dolichoderinae & D & $-$0.991 &  8 & 0.137 & [0.009, 0.481] & 23.3\\
\textit{Myrmecia pyriformis}          & Myrmeciinae    & C &  $+$1.247 &  6 & 0.118 & [0.000, 0.504] & 23.7\\
\textit{Iridomyrmex splendens}        & Dolichoderinae & D & $-$0.703 &  5 & 0.114 & [0.000, 0.495] & 22.8\\
\textit{Rhytidoponera victoriae}      & Ectatomminae   & M & $-$0.110 &  6 & 0.112 & [0.000, 0.494] & 22.9\\
\textit{Melophorus turneri}           & Formicinae     & D & $-$0.996 &  8 & 0.103 & [0.001, 0.472] & 23.6\\
\midrule
\multicolumn{3}{l}{\textbf{Assemblage mean ($\pm$ SD)}} &
  & & $0.21 \pm 0.09$ & &\\
\bottomrule
\end{tabular}
\begin{tablenotes}
\footnotesize
\item[] $\dagger$ Excluded from cross-correlation ($< 20$ observation days);
  $r_{\rm peak}$ and lag: see Table~\ref{tab:xcorr}.
\end{tablenotes}
\end{threeparttable}
\end{sidewaystable}

\begin{table}[!htb]
\caption{H1a evidence classification and assemblage test (Claim~1).  Evidence
  level based on lower bound of 95\,\% Bayesian CrI: Strong ($> 0.05$),
  Moderate ($> 0.01$), Weak ($> 0.00$), None (CrI includes 0).
  \label{tab:h1a_test}}
\begin{threeparttable}
\begin{tabular}{lrrl}
\toprule
\textbf{Evidence level} & \textbf{\textit{N} species} &
\textbf{ICC range} & \textbf{Criterion}\\
\midrule
Strong (CrI$_{\rm lo} > 0.05$)    & 1  & 0.37        & CrI clearly excludes 0\\
Moderate (CrI$_{\rm lo} > 0.01$)  & 8  & 0.16--0.40  & CrI excludes 0\\
Weak (CrI$_{\rm lo} > 0.00$)      & 9  & 0.10--0.40  & CrI lower tail near 0\\
None (CrI includes 0)             & 0  & ---         & Independence plausible\\
\midrule
\textbf{Total} & \textbf{18} & & \\
\midrule
\multicolumn{4}{l}{\textit{Technical check: Wilcoxon signed-rank test}
  \textit{(ICC posterior means vs.\ null of 0; expected significant under Bayesian shrinkage, see note\,b)}}\\
$W = 0$ & $p < 0.001$ & $r_{\rm effect} = 1.00$ &
  All 18 ICC means $> 0$\\
\bottomrule
\end{tabular}
\begin{tablenotes}
\footnotesize
\item[] Effect size $r = W / (n(n+1)/2)$ following \citet{Kerby2014}.
\item[] The Wilcoxon test on Bayesian shrinkage estimates of a non-negative
  quantity is expected to be significant irrespective of true ICC magnitude;
  it is reported for completeness only.  The per-species credible interval
  evidence (rows above) provides the primary assessment of synchrony strength.
\end{tablenotes}
\end{threeparttable}
\end{table}

\subsection{Claim 2: Body size predicts longevity, not synchrony}

\begin{figure}[!htb]
\centering
\includegraphics[width=0.65\textwidth]{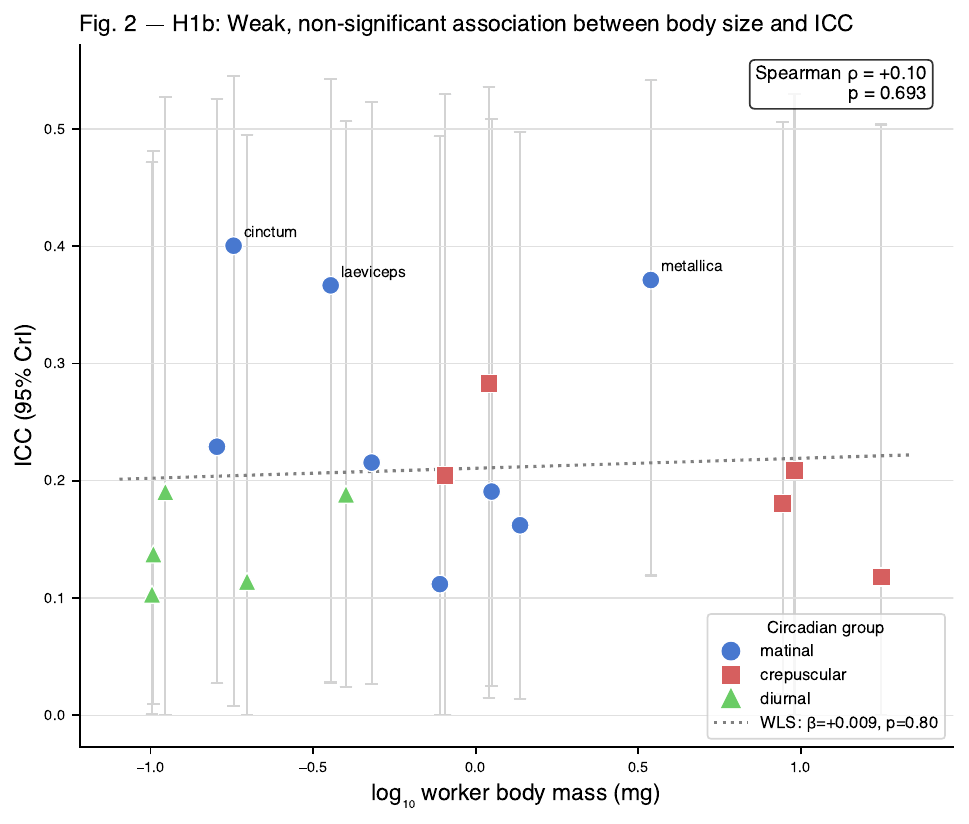}
\caption{\textbf{ICC posterior mean vs.\ log\textsubscript{10} worker body
  mass across 18 species (Claim~2, H1b).}  Vertical bars are 95\,\% Bayesian
  CrI; points are encoded by both colour and shape by circadian activity
  group (matinal\,=\,blue circle; crepuscular\,=\,red square;
  diurnal\,=\,green triangle).  The dashed regression line is the WLS fit
  (simple model: $\beta = +0.009$, $p = 0.80$, $R^2_{\rm adj} = -0.06$).
  The near-flat line and the two extreme-ICC species at opposite ends of the
  mass axis (\textit{Chelaner cinctum} at $\log_{10}\text{mass} = -0.745$,
  ICC\,=\,0.400; \textit{Rhytidoponera metallica} at $+0.539$,
  ICC\,=\,0.371) illustrate the weak, non-significant association between
  body size and synchrony (Spearman $\rho = +0.10$, $p = 0.69$).
  \label{fig:icc_vs_size}}
\end{figure}

The trait that predicts mean longevity does not predict synchrony.  Across
species, body mass was weakly \emph{positively} correlated with ICC ---
the opposite of the buffering prediction --- and far from significant
(Spearman $\rho = +0.10$, $p = 0.69$; Fig.~\ref{fig:icc_vs_size}).
Under both WLS specifications the body-size coefficient was small, positive,
and non-significant (Table~\ref{tab:wls}); the 95\,\%~CI from the simple
model ($[-0.061,\,+0.078]$) and the phylogenetically corrected PGLS
($[-0.072,\,+0.069]$; Pagel's $\lambda = 0.12$, $p = 0.79$) together
exclude any effect of biologically relevant magnitude under either framework.
The decoupling is made vivid by the two highest-ICC species, which bracket
almost the entire mass range: the small \textit{Chelaner cinctum}
(log\textsubscript{10} mass\,=\,$-0.745$, ICC\,=\,0.400) and the large
\textit{Rhytidoponera metallica} (log\textsubscript{10} mass\,=\,$+0.539$,
ICC\,=\,0.371).

The null was robust to restricting the analysis to the nine species with
the most precise ICC estimates (CrI$_{\rm lo} > 0.01$; Moderate or Strong
evidence): Spearman $\rho = -0.30$, $p = 0.43$; WLS $\beta = +0.005$,
$p = 0.93$ ($N = 9$).  The direction was marginally consistent with
buffering in this precise subset, but far from significant and based on
nine data points.

\begin{table}[!htb]
\caption{WLS regression of ICC on body size with and without subfamily
  control (H1b, Claim~2).  Weights\,=\,$1 / \text{CrI\_width}^2$.
  Reference: Dolichoderinae.  $N = 18$ species.
  \label{tab:wls}}
\begin{threeparttable}
\textbf{Model 1: ICC\,$\sim$\,body size\,+\,subfamily
  ($R^2_{\rm adj} = +0.06$)}\\[4pt]
\begin{tabular}{lrrrrl}
\toprule
\textbf{Term} & $\boldsymbol{\beta}$ & \textbf{SE} & $\boldsymbol{t}$ &
\textbf{95\,\% CI} & \textbf{\textit{p}}\\
\midrule
Intercept & $+$0.194 & 0.048 & 4.063 & [0.090, 0.298] & 0.002\\
Body size (log\textsubscript{10} mass, centred) &
  $+$0.049 & 0.050 & 0.980 & [$-$0.060, 0.159] & 0.346\\
Subfamily: Ectatomminae  & $+$0.019 & 0.077 & 0.241 & [$-$0.150, 0.187] & 0.814\\
Subfamily: Formicinae    & $-$0.012 & 0.068 & $-$0.169 & [$-$0.160, 0.137] & 0.868\\
Subfamily: Myrmeciinae   & $-$0.096 & 0.119 & $-$0.806 & [$-$0.354, 0.163] & 0.436\\
Subfamily: Myrmicinae    & $+$0.112 & 0.062 & 1.820 & [$-$0.022, 0.246] & 0.094\\
\bottomrule
\end{tabular}

\medskip

\textbf{Model 2: ICC\,$\sim$\,body size only ($R^2_{\rm adj} = -0.06$)}\\[4pt]
\begin{tabular}{lrrrrl}
\toprule
\textbf{Term} & $\boldsymbol{\beta}$ & \textbf{SE} & $\boldsymbol{t}$ &
\textbf{95\,\% CI} & \textbf{\textit{p}}\\
\midrule
Intercept & $+$0.209 & 0.022 & 9.375 & [0.162, 0.257] & $<$0.001\\
Body size (log\textsubscript{10} mass, centred) &
  $+$0.009 & 0.033 & 0.260 & [$-$0.061, 0.078] & 0.798\\
\bottomrule
\end{tabular}
\begin{tablenotes}
\footnotesize
\item[] Both models find no evidence that body size buffers synchrony: the
  body-size coefficient is small, positive (opposite to the buffering
  prediction), and far from significant under either specification
  (Spearman $\rho = +0.10$, $p = 0.69$).  Body size: log\textsubscript{10}
  worker body mass (mg), grand-mean centred.  Model~1 residual
  df\,=\,12; Model~2 residual df\,=\,16.
\end{tablenotes}
\end{threeparttable}
\end{table}

Broadening the comparison to subfamily level, the relationship is
directionally consistent with buffering but statistically inconclusive
(Spearman $\rho = -0.30$, $p = 0.62$, $N = 5$;
Table~S1 (Supplementary Material)): the smallest-bodied subfamily,
Myrmicinae, had the highest mean ICC (0.297) and the largest-bodied,
Myrmeciinae, the lowest (0.163), but with only five subfamilies the test is
severely underpowered.

\subsection{Claim 3: Temperature is a suggestive, lagged correlate}

\begin{figure}[!htb]
\centering
\includegraphics[width=0.65\textwidth]{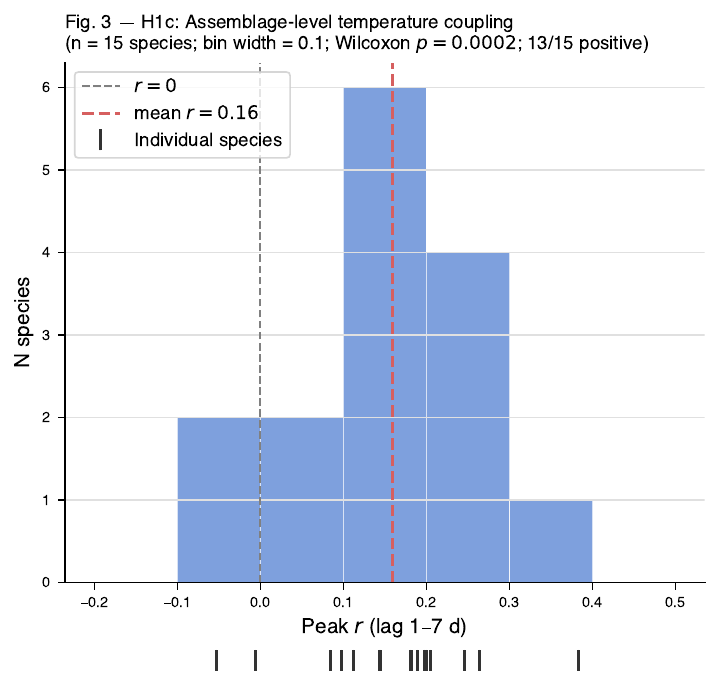}
\caption{\textbf{Distribution of peak cross-correlations $r_{\rm peak}$ across
  15 species (Claim~3, H1c).}  Histogram (bin width\,=\,0.10) of the peak
  Pearson correlation between species-level daily mortality and maximum
  temperature within the 1--7 day lag window.  Tick marks below the $x$-axis
  (rug plot) show the individual species values.  The vertical dashed grey
  line marks $r = 0$; the red dashed line marks the assemblage mean
  ($r = 0.16$).  Thirteen of 15 species show a positive peak correlation
  (Wilcoxon $W = 117$, $p = 0.0002$).  Because each value is the maximum
  across seven lags, peak correlations are upwardly biased relative to a
  window-averaged summary (Table~S2).
  \label{fig:rpeak_hist}}
\end{figure}

\begin{figure}[!htb]
\centering
\includegraphics[width=\textwidth]{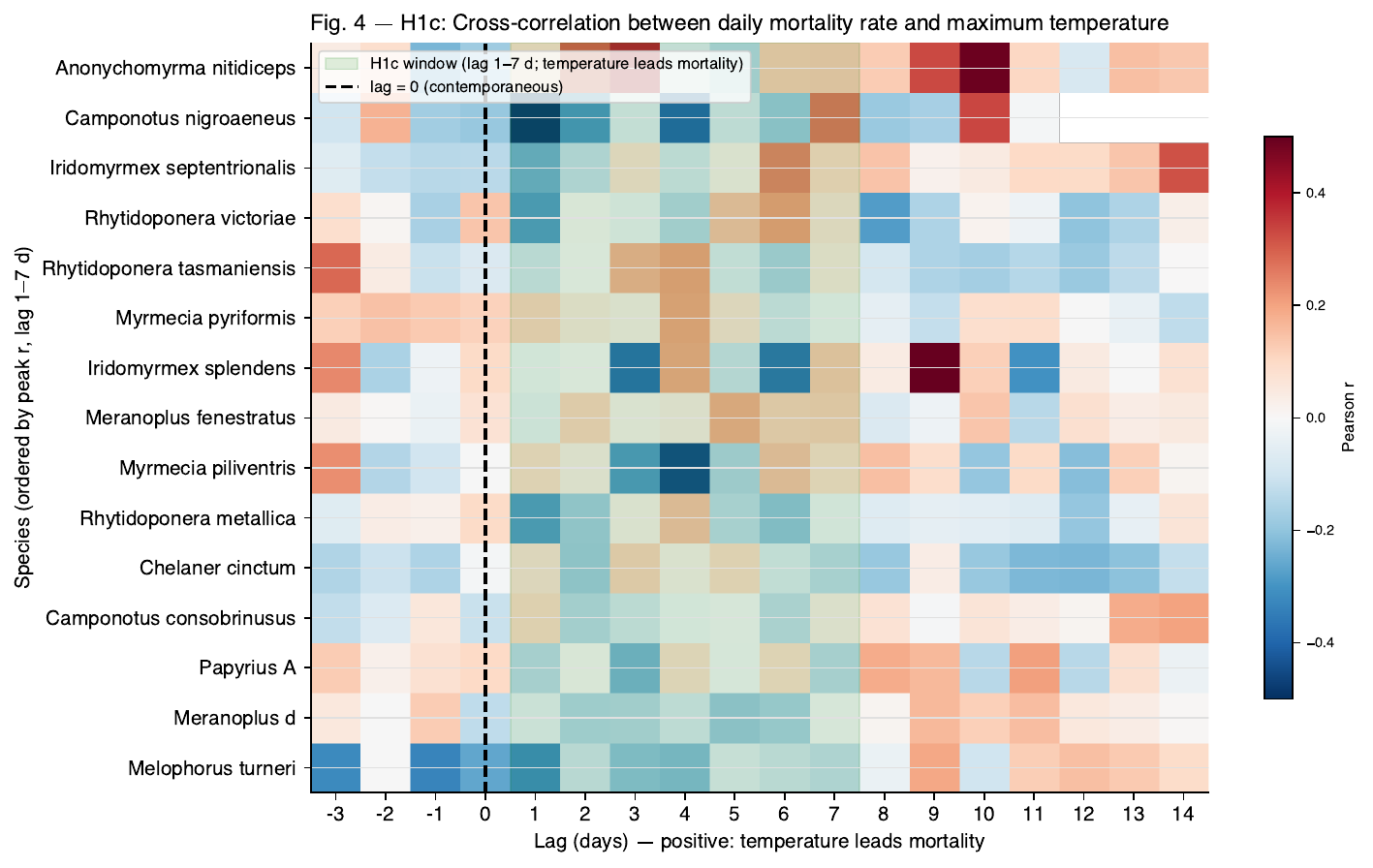}
\caption{\textbf{Cross-correlation heatmap: species-level daily mortality vs.\
  maximum temperature at lags $-3$ to $+14$ days (Claim~3, H1c).}
  $N = 15$ species with $\geq 20$ observation days.  Rows ordered by peak
  cross-correlation within the 1--7 day window (highest at top); columns are
  lags.  Colour scale: Pearson $r$ (warm\,=\,positive; cool\,=\,negative).
  The shaded vertical zone marks the 1--7 day window of interest; the dashed
  line marks lag 0 (contemporaneous).  The positive cluster within the shaded
  zone (present in 13/15 species) is consistent with a suggestive lagged
  temperature--mortality signal detectable at assemblage level only; no
  individual species survived Bonferroni correction
  ($\alpha_{\rm adj} = 0.0033$).
  \label{fig:xcorr_heatmap}}
\end{figure}

Heat events leave a demographic signature, but a modest one.  Across the 15
analysable species, daily mortality cross-correlated positively with maximum
temperature within a 1--7 day window (mean peak $r = 0.16 \pm 0.11$ SD;
Wilcoxon $W = 117$, $p = 0.0002$, $r_{\rm effect} = 0.83$;
Table~\ref{tab:h1c_assemblage}, Fig.~\ref{fig:xcorr_heatmap},
Fig.~\ref{fig:rpeak_hist}), with 13 of 15
species showing a positive correlation at their best lag in that window.  The
effect is consistent in direction but small in magnitude: no species
yielded a cross-correlation that survived Bonferroni correction for 15 tests
($\alpha_{\rm adj} = 0.0033$), and the nominally strongest individual signal
was \textit{Anonychomyrma nitidiceps} (lag = 3 days, $r = 0.383$, $p = 0.006$,
uncorrected; $p_{\rm PP} = 0.013$ after Pyper--Peterman autocorrelation
correction, still above the Bonferroni threshold).  We therefore interpret
the temperature--mortality coupling as a suggestive assemblage-level pattern
rather than a confirmed within-species signal.  This interpretation is
reinforced by a sensitivity analysis: when we replaced the peak correlation
with the unweighted mean correlation across all lags in the 1--7 day window
--- a summary that does not capitalise on the single best lag --- the positive
signal disappeared (mean $r = -0.03 \pm 0.08$ SD; Wilcoxon $p = 0.93$;
only 4 of 15 species positive; Table~S2 (Supplementary Material)).
The coupling is thus concentrated at a species-specific lag rather than
spread across the window, and the assemblage-level result is best read as
evidence of a consistent \emph{direction} at the best lag, not of a pervasive
thermal effect.

\begin{figure}[!htb]
\centering
\includegraphics[width=\textwidth]{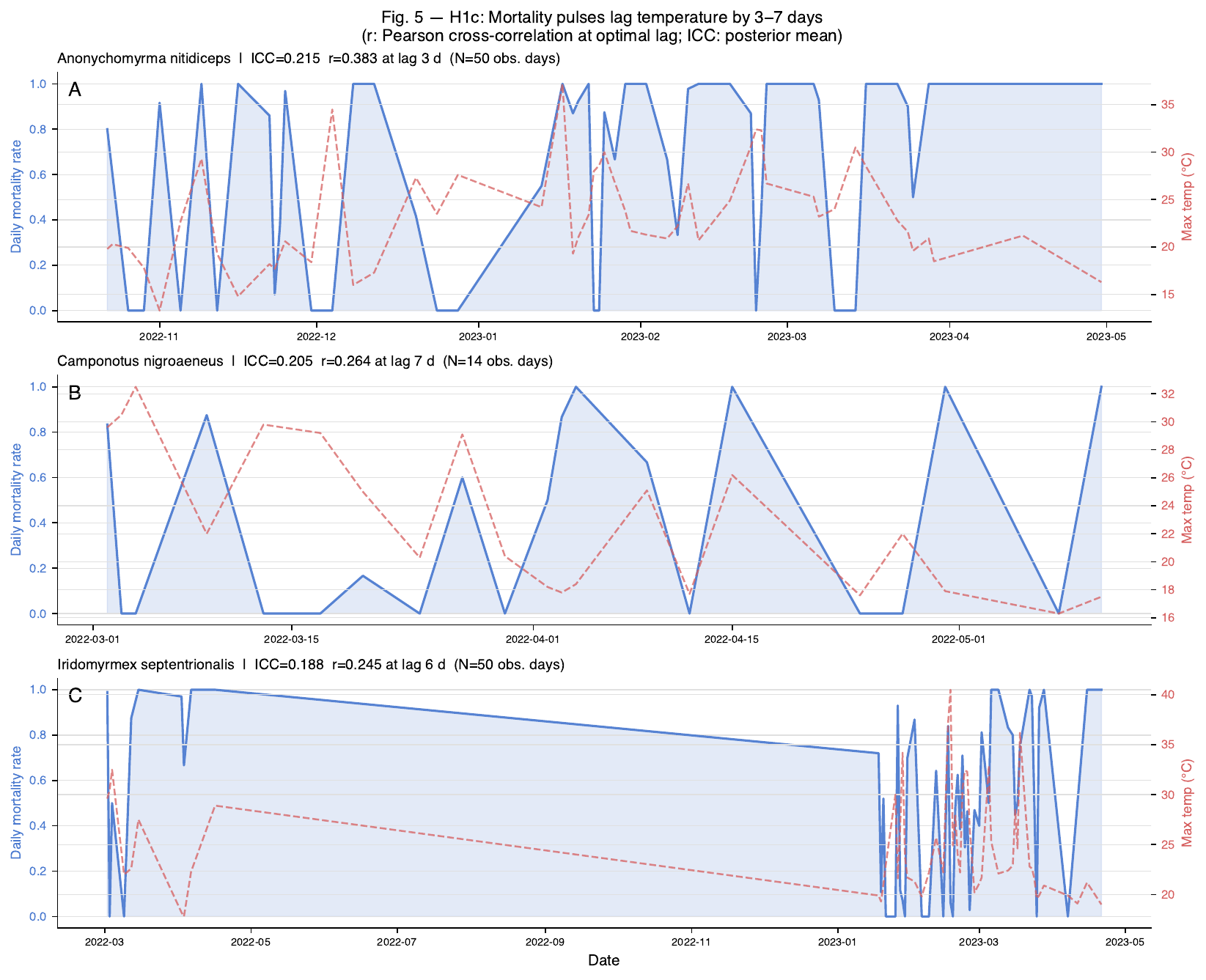}
\caption{\textbf{Dual-axis time series for the three most
  temperature-responsive species (Claim~3, H1c).}  Panels A--C show the three
  species with the highest peak cross-correlation.  Each panel shows
  species-level daily mortality rate (left axis, blue line) and daily
  maximum temperature (right axis, red dashed line).  Panel titles report the
  species name, ICC posterior mean, peak Pearson cross-correlation $r$ with
  its optimal lag, and the number of observation days ($N$).  The rightward
  offset of mortality peaks relative to temperature peaks illustrates the
  3--7 day physiological lag.
  \label{fig:time_series}}
\end{figure}

The next strongest couplings were
\textit{Camponotus nigroaeneus} (lag = 7 d, $r = 0.264$),
\textit{Iridomyrmex septentrionalis} (lag = 6 d, $r = 0.245$),
\textit{Rhytidoponera victoriae} (lag = 6 d, $r = 0.205$), and
\textit{Rhytidoponera tasmaniensis} (lag = 4 d, $r = 0.200$;
Table~\ref{tab:xcorr}, Fig.~\ref{fig:time_series}).  The characteristic 3--7
day delay is consistent with a delayed physiological response rather than
instantaneous heat death \citep{Colinet2015, Sales2021}.

Crucially, temperature coupling does not explain synchrony.  ICC was
unrelated to the strength of temperature coupling (Spearman $\rho = -0.004$,
$p = 0.99$): the two highest-ICC species show only moderate coupling, and the
most temperature-coupled species (\textit{A. nitidiceps}) has a middling
ICC\,=\,0.215.

\begin{table}[!htb]
\caption{H1c assemblage-level test: Wilcoxon signed-rank test on peak
  cross-correlations $r$ at lag 1--7 days (one-sided, greater; Claim~3).
  \label{tab:h1c_assemblage}}
\begin{threeparttable}
\begin{tabular}{lrrrrl}
\toprule
\textbf{Test} & $\boldsymbol{N}$ & \textbf{Mean} $\boldsymbol{r}$ &
$\boldsymbol{W}$ & \textbf{\textit{p}} & \textbf{Support}\\
\midrule
H1c: peak $r$ at lag 1--7 d $> 0$ &
  15 & $0.16 \pm 0.11$ (SD) & 117 & 0.0002 & \checkmark\\
\midrule
\multicolumn{6}{l}{13/15 species positive at best lag; effect size
  $r_{\rm W} = 0.83$}\\
\multicolumn{6}{l}{ICC vs.\ $r_{\rm peak}$ (Spearman): $\rho = -0.004$,
  $p = 0.99$ (null)}\\
\multicolumn{6}{l}{Excluded: \textit{C. claripesA} ($n = 18$),
  \textit{C. laeviceps} ($n = 9$), \textit{I. notialis} ($n = 15$)}\\
\bottomrule
\end{tabular}
\end{threeparttable}
\end{table}

\begin{table}[!htb]
\caption{Pearson cross-correlation between species-level daily mortality and
  maximum temperature, at lag 1--7 days (H1c, Claim~3; $N = 15$ species with
  $\geq 20$ observation days, ordered by peak $r$).  $p$: uncorrected
  $p$-value at peak lag.
  \label{tab:xcorr}}
\begin{threeparttable}
\small
\begin{tabular}{llrrrrr}
\toprule
\textbf{Species} & \textbf{Subfamily} &
$\boldsymbol{r_{\rm peak}}$ & \textbf{Lag (d)} &
\textbf{\textit{p}} & \textbf{ICC} & \textbf{\textit{N} obs.}\\
\midrule
\textit{Anonychomyrma nitidiceps}    & Dolichoderinae & 0.383 & 3 & 0.006$^{\star}$ & 0.215 & 50\\
\textit{Camponotus nigroaeneus}      & Formicinae     & 0.264 & 7 & 0.362 & 0.205 & 14\\
\textit{Iridomyrmex septentrionalis} & Dolichoderinae & 0.245 & 6 & 0.086 & 0.188 & 50\\
\textit{Rhytidoponera victoriae}     & Ectatomminae   & 0.205 & 6 & 0.167 & 0.112 & 47\\
\textit{Rhytidoponera tasmaniensis}  & Ectatomminae   & 0.200 & 4 & 0.136 & 0.162 & 57\\
\textit{Myrmecia pyriformis}         & Myrmeciinae    & 0.198 & 4 & 0.109 & 0.118 & 67\\
\textit{Iridomyrmex splendens}       & Dolichoderinae & 0.190 & 4 & 0.325 & 0.114 & 29\\
\textit{Meranoplus fenestratus}      & Myrmicinae     & 0.181 & 5 & 0.080 & 0.191 & 94\\
\textit{Myrmecia piliventris}        & Myrmeciinae    & 0.145 & 6 & 0.543 & 0.209 & 20\\
\textit{Rhytidoponera metallica}     & Ectatomminae   & 0.143 & 4 & 0.158 & 0.371 & 99\\
\textit{Chelaner cinctum}            & Myrmicinae     & 0.112 & 3 & 0.216 & 0.400 & 123\\
\textit{Camponotus consobrinusus}    & Formicinae     & 0.097 & 1 & 0.357 & 0.181 & 92\\
\textit{Papyrius} A                  & Dolichoderinae & 0.084 & 6 & 0.529 & 0.137 & 58\\
\textit{Meranoplus d}                & Myrmicinae     & $-$0.006 & 7 & 0.968 & 0.229 & 51\\
\textit{Melophorus turneri}          & Formicinae     & $-$0.053 & 5 & 0.723 & 0.103 & 48\\
\midrule
\textbf{Mean $\pm$ SD} & & $0.16 \pm 0.11$ & & & &\\
\bottomrule
\end{tabular}
\begin{tablenotes}
\footnotesize
\item[] $\star$ Nominally lowest $p$ (uncorrected); does not survive Bonferroni
  correction ($\alpha_{\rm adj} = 0.0033$).
  Spearman $\rho$(ICC vs.\ $r_{\rm peak}$)\,=\,$-0.004$, $p = 0.99$.
\item[] Pyper--Peterman autocorrelation-corrected $p$-values are reported in
  the supplementary data files.  The nominally strongest signal
  (\textit{A.~nitidiceps}, $p = 0.006$) becomes $p_{\rm PP} = 0.013$ after
  correction; no species survives Bonferroni under either $p$-value.
\end{tablenotes}
\end{threeparttable}
\end{table}

\section{Discussion}
\label{sec:discussion}

Worker mortality in this diverse ant assemblage is neither random nor
independent: cohorts die in pulses, the pulse is partly but not fully
thermal, and body size --- despite setting how long workers live --- has no
bearing on \emph{when} they die together.  Synchrony is real and widespread
(Claim~1); body size is silent on it (Claim~2); and temperature contributes
but does not account for it (Claim~3).  What unifies all three results is
that \textbf{the magnitude of mortality synchrony is governed by something
other than the variables that govern mean longevity or that produce the
clearest acute mortality signal.}

\subsection{Synchrony is a real, overlooked dimension of worker life history}

Half the assemblage had credible intervals excluding zero (9/18 species);
no species had a posterior mean near zero (minimum ICC\,=\,0.103).
A shared component accounting for ${\sim}21\,\%$ of daily mortality variance
is substantial, and it has consequences beyond description.  Colony fitness
depends not only on \emph{how long} workers live but on \emph{when} they die:
synchronised mortality creates bottlenecks in worker availability that
independent-survival models --- the standard in colony demography --- cannot
capture, consistent with a systematic overestimation of colony resilience to
environmental stress --- a hypothesis that warrants formal testing with
colony-level demographic data.

To put this concretely: with ICC\,=\,0.21, the variance in deaths within a
typical cohort is inflated approximately fivefold relative to what
independent survival would predict \citep[design effect;][]{Nakagawa2026}.
Episodic shortfalls severe enough to impair brood care or colony defence
would therefore be expected to occur far more often than standard
colony-demography models would predict --- and are particularly likely
during the heat events identified in Claim~3.

\subsection{Body size predicts duration but not synchrony (Claim 2)}

The decoupling between body size and ICC is the conceptual core of the
paper: in this assemblage body size strongly predicts mean worker longevity
\citep{Riskas2026} yet is silent on synchrony --- a powerful predictor of
\emph{one} mortality axis and uninformative about \emph{another}.
Mechanistically, this suggests that the advantages of large size --- lower
metabolic rate, greater antioxidant capacity, larger reserves --- act on the
\emph{baseline hazard} of individual ageing but appear not to buffer a
colony's \emph{susceptibility to acute, shared shocks} --- an inference from
the ICC--size decoupling observed here, not a direct claim of the cited works.
When an extrinsic force hits, large and small workers appear to die at
similarly elevated and correlated rates.  This points to acute thermal
tolerance --- e.g.\ critical thermal maximum ($\text{CT}_{\rm max}$) --- as a
potentially more informative predictor of ICC than body mass
\citep{Kaspari2015, Roeder2021}, a hypothesis directly testable in future
trait-based work.

\subsection{Temperature: a lagged correlate, but not sufficient to explain synchrony
  (Claim 3)}

The positive temperature--mortality cross-correlation at 1--7 days provides
suggestive evidence that heat events are associated with synchronised mortality
pulses in a diverse assemblage, with a delay consistent with a
cumulative-damage cascade \citep{Colinet2015, Sales2021}.  But two features
temper the story.  First, the effect is modest (mean $r \approx 0.16$) and no
species crosses the Bonferroni threshold for individual significance
($\alpha_{\rm adj} = 0.0033$), so the coupling is detectable only at the
assemblage level.

Second, and more tellingly, ICC is uncorrelated with temperature coupling
($\rho = -0.004$): the most synchronised species are not the most
temperature-tracked.  Synchrony must therefore have additional correlates ---
rainfall and foraging suppression, pathogen dynamics, or collective
thermoregulation failures during extremes --- that our single-covariate
analysis cannot resolve.

A counter-intuitive detail reinforces the role of thermal physiology over
simple exposure: the most temperature-coupled species, \textit{A. nitidiceps},
is a small matinal forager presumably active before peak heat.  Its strong
coupling may reflect low thermal tolerance or exposed microhabitat use rather
than midday heat exposure per se --- again implicating species-specific
tolerance over body size.

\subsection{Why this matters: from worker pulses to colony and community}

Three practical consequences follow. First, \textbf{demographic risk may
be systematically mis-estimated.} Colony-demography and population-viability
models that treat worker deaths as independent would understate the variance in
worker availability roughly fivefold at the observed ICC\,=\,0.21
\citep[design effect;][]{Nakagawa2026}. Such models would therefore underestimate
how often a colony falls below the worker threshold needed for brood care or
defence, potentially biasing assessments of which species and colonies are most
at risk of episodic collapse.

Second, \textbf{the size--synchrony decoupling undermines a common shortcut.}
Body mass is inexpensive to measure and reliably ranks species by mean
longevity \citep{Riskas2026}, making it an attractive trait for screening
demographic vulnerability. Our results show it carries no information about
\emph{synchrony}: a vulnerability assessment built on body size alone would
miss the temporal-clustering axis entirely. Predicting which species suffer
the sharpest mortality pulses will require physiological traits --- acute
thermal tolerance ($\text{CT}_{\rm max}$), microhabitat exposure
\citep{Kaspari2015, Roeder2021} --- not body size.

Third, \textbf{the thermal signature links worker synchrony to thermal
extremes.} Because mortality pulses tracked maximum temperature at a 3--7 day
lag, any increase in the frequency or intensity of extreme-heat events would,
through the same mechanism, translate into more frequent and more severe
synchronised worker shortfalls. And because synchrony concentrates deaths in
time, its community-level footprint is episodic rather than gradual: the
ecosystem functions ants provide --- seed dispersal, soil turnover, predation
--- could tentatively fail in pulses tied to heat events rather than declining
smoothly, though this inference extends beyond the present correlational
dataset.

Worker-mortality synchrony is thus a candidate mechanism by which thermal
extremes propagate from individual physiology to colony persistence and
community function.

\subsection{Synthesis and future directions}

Mortality synchrony is widespread, size-independent, and only partly thermal.
The natural next step is trait-based: measuring $\text{CT}_{\rm max}$ and
microhabitat thermal exposure across species to test whether acute thermal
tolerance predicts ICC where body mass fails, and assembling multi-year,
pathogen- and microclimate-resolved datasets to partition the non-temperature
component of synchrony.  No published multi-species $\text{CT}_{\rm max}$
dataset currently covers a sufficient fraction of the 18 Australian genera
studied here \citep{Roeder2021}; compiling or directly measuring
$\text{CT}_{\rm max}$ for this assemblage is therefore the single
highest-priority empirical extension of the present work.

\subsection{Limitations}

\begin{enumerate}
  \item \textbf{ICC precision and power.}  ICC estimates come from the hierarchical
    daily-survival model of \citet{Riskas2026}, which was designed to estimate
    baseline hazard rates, not to maximise ICC precision; credible intervals are
    accordingly wide (mean CrI width\,$\approx$\,0.50), reflecting few cohorts
    per species (median $\approx$\,5) rather than a modelling inadequacy.
    Given the observed standard error of the body-size coefficient
    (SE\,=\,0.033 in Model~2), the WLS regression had 80\,\% power to detect
    associations larger than $|\beta| \approx 0.09$ ICC units per
    $\log_{10}$\,mg (equivalent to a ${\sim}0.20$ ICC change across the full
    body-mass range).  The absence of any signal at this sensitivity argues
    against a biologically relevant buffering effect, and the sensitivity
    analysis restricted to the nine most precise species
    (CrI$_{\rm lo} > 0.01$) confirmed the same null
    (Spearman $\rho = -0.30$, $p = 0.43$; WLS $\beta = +0.005$, $p = 0.93$).

  \item \textbf{Short series for some species.}  Three species were excluded
    from the cross-correlation for $< 20$ observation days, so species with
    sparse temporal coverage are under-represented in the temperature--coupling
    analysis; if these species differ systematically in their mortality
    dynamics, the assemblage-level coupling estimate could be affected.

  \item \textbf{Single thermal covariate.}  We used station maximum air
    temperature; operative, soil, and shaded-microhabitat temperatures would
    be more physiologically relevant but are unavailable at the needed
    resolution.  Because maximum air temperature is an imperfect proxy for the
    thermal stress experienced inside nests, it may under- or over-represent
    true thermal exposure, which could attenuate the observed
    temperature--mortality correlation (if actual temperatures are higher than
    recorded) or inflate it (if foragers are preferentially exposed to extreme
    ambient temperatures).

  \item \textbf{Observational design.}  All results are correlational; causal
    inference would require experimental thermal manipulation or heatwave
    quasi-experiments.  Multiple co-varying stressors could independently or
    jointly drive synchrony: rainfall events, reduced foraging activity during
    heatwaves, seasonal disease dynamics, and collective thermoregulation
    failures are plausible alternative or complementary explanations that our
    single-covariate design cannot disentangle.

  \item \textbf{Statistical caveats.}  Both the mortality-rate and temperature
    time series carry positive autocorrelation.  We corrected per-species
    Pearson $p$-values using the Pyper--Peterman effective sample size
    \citep{PyperPeterman1998}; the correction reduces individual-species
    significance but does not affect the assemblage-level Wilcoxon test, which
    tests consistent direction rather than individual magnitude.  The
    species-level daily mortality rate also has a small denominator on many
    days (median ${\approx}5$ active cohorts), producing the extreme 0 and 1
    values visible in Fig.~\ref{fig:time_series}; count-based or binomial models
    would provide more precise coupling estimates in future work.  Finally,
    because the assemblage-level temperature test selects the peak correlation
    within a seven-lag window, it is upwardly biased under the null; the
    sensitivity analysis using the window-averaged correlation (Table~S2)
    confirms that the positive signal is specific to each species' best lag
    and does not hold as a broad, window-wide association.  The
    temperature--mortality coupling should accordingly be regarded as a
    directional hint requiring confirmation, not an established driver.

  \item \textbf{Phylogenetic non-independence.}  The assemblage includes
    multiple congeneric pairs (\textit{Rhytidoponera}, \textit{Iridomyrmex},
    \textit{Camponotus}, \textit{Myrmecia}, \textit{Meranoplus}).  PGLS with
    Pagel's $\lambda$ estimated by maximum likelihood yielded
    $\lambda = 0.12$ (LRT $p = 0.79$), indicating that ICC shows negligible
    phylogenetic signal across this assemblage.  The OLS/WLS analyses are
    therefore appropriate, and the size--synchrony null is not an artefact
    of phylogenetic non-independence (see Results, Claim~2).
\end{enumerate}

\subsection{Conclusions}

Within-species worker mortality in Australian ants is synchronised across the
assemblage (mean ICC\,=\,0.21 across 18 species), driven by shared extrinsic
forces rather than individual ageing variation.  Body size --- despite
predicting mean longevity --- does not predict synchrony, decoupling
\emph{how long} from \emph{when} workers die.  Temperature pulses precede
mortality pulses by 3--7 days across the assemblage but explain neither which
species are most synchronised nor most of the synchrony itself.
Species-specific thermal tolerance, not body mass, is the more promising
candidate for completing the mechanistic link from physiology to colony-level
mortality dynamics.

Recognising synchrony as a distinct axis of worker life history matters in
practice: demographic models that ignore it would be expected to understate
how often colonies face acute worker shortfalls, and --- because these
shortfalls track heat --- this blind spot is likely to widen as thermal
extremes intensify.

\section*{Data and Code Availability}

All data are from \citet[][doi: 10.5061/dryad.j0zpc86s4]{Riskas2026}.
Analysis scripts are openly available at
\url{https://github.com/rafa-rodriguess/australian\_ants\_how\_long\_versus\_when\_worker\_mortalit}.

\section*{Conflict of Interest}

The authors declare no competing interests.

\bibliography{paperB}

\end{document}